\documentclass[prd,twocolumn,showpacs,a4paper]{revtex4}
\begin{document}
\title{Holography and the Large Number Hypothesis}
\author{Guillermo A. Mena Marug\'{a}n$^1$ and Saulo Carneiro$^{1,2}$}
\address{$^1$ I.M.A.F.F.,
C.S.I.C., Serrano 121, 28006 Madrid, Spain\\$^2$ Instituto de
F\'{\i}sica, Universidade Federal da Bahia, 40210-340, Salvador, Bahia,
Brazil}

\begin{abstract}
Dirac's large number hypothesis is motivated by certain scaling
transformations that relate the parameters of macro and
microphysics. We show that these relations can actually be
explained in terms of the holographic $N$ bound conjectured by
Bousso and a series of purely cosmological observations, namely,
that our universe is spatially homogeneous, isotropic, and flat to
a high degree of approximation and that the cosmological constant
dominates the energy density at present.
\end{abstract}
\maketitle

Explaining the value of the constants of nature is one of the most
exciting challenges of theoretical physics. Some of these
constants play a fundamental role in the foundations of the
scientific paradigms. This is the case of Planck constant $\hbar$
in quantum mechanics, and of Newton constant $G$ and the speed of
light $c$ in general relativity. These three constants provide a
natural system of units for all physical quantities. For instance,
the length and mass units are $l_P=\sqrt{\hbar
G/c^{3}}=1.6\times10^{-35}$ m and $m_P=\sqrt{\hbar c/G}= 2.2\times
10^{-8}$ kg. In terms of these Planck units, the other constants
of nature become dimensionless numbers.

Already in the 20's, Eddington tried unsuccessfully to deduce the
value of all constants of physics from theoretical considerations
\cite{Ed}. Most importantly, he pointed out the existence of
relations between the parameters of fields that at first sight
seem unconnected, like nuclear physics and cosmology. Among these,
perhaps the most intriguing relation is the apparent coincidence
between the present number of baryons in the universe, known as
Eddington number, and the squared ratio of the electric to the
gravitational force between the proton and the electron. This
coincidence between large numbers can also be expressed in the
alternative form \cite{Weinberg}
\begin{equation}\label{rel} \hbar^2H_0\approx
Gcm_N^3.\end{equation} This approximate identity is sometimes
called the Eddington-Weinberg relation. Here, $m_N$ is the proton
mass and $H_0\approx 70$ km/(sMpc) is the present value of the
Hubble constant \cite{Hubble}.

Actually, the Hubble parameter is not a true constant, but varies
as the inverse of the cosmological time in standard
Friedman-Robertson-Walker (FRW) cosmology \cite{Weinberg}. This
fact led Dirac \cite{Dirac} to put forward the hypothesis that
Newton constant must depend on time as $H_0$, $G\propto t^{-1}$,
so that relation (1) is always valid. In spite of its attractive
features, Dirac's large number hypothesis turns out to be
incompatible with the experimental bounds that exist on the time
variation of $G$ \cite{sol}. Therefore, the explanation of the
Eddington-Weinberg relation still remains a mystery.

Recently, the determination of cosmological parameters has
experienced a considerable revolution. The observation of type Ia
supernovae (SNe Ia) at high redshift has provided evidence in
favor of a positive cosmological constant \cite{sne}. In addition,
accurate measurements of the angular power spectrum of
anisotropies in the cosmic microwave background (CMB) have shown
that the curvature of the universe is close to flat \cite{CMB}.
These CMB and SNe Ia data, together with other cosmological
information, have been combined in a consistent (nearly) flat FRW
model whose values of the cosmological constant $\Lambda$ and
matter density $\rho_0$ are, approximately, $c^2 \Lambda= 16 \pi G
\rho_0= 2 H_0^2$ \cite{conc}.

This value of the cosmological constant poses two puzzles. On the
one hand, one would expect that $\Lambda$ emerged from vacuum
fluctuations. In a theory of quantum gravity, these fluctuations
would have Planck energy density. The discrepancy from this
theoretical expectations is of nearly 120 orders of magnitude,
since, in Planck units, $H_0\approx 10^{-60}$. This is the
so-called cosmological constant problem \cite{ccp}. The value of
$\Lambda$, on the other hand, is constant, whereas the density of
matter decreases with expansion. As a consequence, the relation
$16\pi G\rho_0\approx c^2 \Lambda$ is not valid in most of the
history of the universe. Why is it precisely now that the matter
content and $\Lambda$ provide similar contributions to the energy?
This additional puzzle is known as the cosmic coincidence problem
\cite{coinc}.

A new perspective of the cosmological constant problem, which puts
the emphasis on fundamental aspects of gravity rather than in
purely quantum field theory (QFT) considerations, has recently
emerged with the advent of holography \cite{holo}. In an
over-simplified version, the holographic principle states that the
entropy $S$ \cite{entro} of a physical system subject to gravity
is bounded from above by a quarter of its boundary area in Planck
units, $S\leq A/(4l_p^{2})$. From this point of view, the physical
degrees of freedom are not proportional to the volume in the
presence of the gravitational field, but reside in the bounding
surface.

A more rigorous, covariant formulation of the holographic
conjecture has been elaborated by Bousso, providing in principle
an entropy bound on null hypersurfaces \cite{bo,Bousso}. Other
less general holographic proposals that find straightforward
application to spatial volumes in cosmology have also been
suggested \cite{FS,BR,H}. In this respect, an issue of debate has
been the largest region of the universe in which an entropy bound
may be feasible. Fischler and Susskind \cite{FS} originally
proposed to consider the particle horizon, at least for adiabatic
evolution, but other possibilities that appear more natural were
soon suggested. One such possibility is the use of the
cosmological apparent horizon, which bounds an anti-trapped region
and has an associated notion of gravitational entropy
\cite{bo,BR}. Another proposal that has found considerable support
is the restriction to the Hubble radius $cH_0^{-1}$ \cite{H},
since this supplies the scale of causal connection beyond which
gravitational perturbations on a flat background cannot grow with
time. It is worth noting, anyway, that for a flat FRW model like
the one that possibly describes our universe, the apparent and
Hubble horizons do in fact coincide \cite{BR}.

For any spacetime with a positive cosmological constant, Bousso
\cite{Nbound} has argued that the holographic principle leads to
the prediction that the number of degrees of freedom $N$ available
in the universe is related to $\Lambda$ by
\begin{equation}\label{Nbound} N=\frac{3\pi}{\Lambda l_P^2\ln{2}}.
\end{equation} The observable entropy $S$ is then bounded by
$N\ln{2}$. This conjecture is called the $N$ bound. Under
quantization, the system would be describable by a Hilbert space
of finite dimension (equal to $2^N$). Bousso's conjecture is
largely influenced by Banks' ideas about the cosmological constant
\cite{Banks}. According to Banks, $\Lambda$ should not be
considered a parameter of the theory; rather, it is determined by
the inverse of the number of degrees of freedom. From this
viewpoint, the cosmological constant problem disappears, because
$N$ can be regarded as part of the data that describe the system
at a fundamental level. Based also on holography, other possible
explanations have been proposed for the value of $\Lambda$ that
are closer in spirit to the standard methods of QFT \cite{CT}.

Since the cosmological constant affects the large scale structure
of the universe but should originate from effective local vacuum
fluctuations, it may provide a natural connection between macro
and microphysics. In addition, $\Lambda$ is related to the number
of degrees of freedom by the holographic principle. As a
consequence, one could expect that holography would play a
fundamental role in explaining the coincidence of the large
numbers arising in cosmology and particle physics. A first
indication that this intuition may work is provided by Zizzi's
work \cite{Zizzi}, who recovered Eddington number starting with a
discrete quantum model for the early universe that saturates the
holographic bound. The main aim of the present paper is to prove
that the large number hypothesis and the holographic conjecture
are in fact not fully independent. To be more precise, we will
show that, in a homogeneous, isotropic, and (quasi)flat universe
like ours, the relations between large numbers can be explained by
the holographic principle assuming that the present energy density
is nearly dominated by $\Lambda$.

The scaling relations that lie behind the large number hypothesis
can be expressed in the form
\begin{eqnarray}\label{lE} l_N&\approx&\Omega l_P,\\
\label{mE} m_N&\approx& \Omega^{-1} m_P,\\ \label{lU} l_U\equiv
cH_0^{-1}&\approx &\Omega^3 l_P,\\ \label{mU} m_U&\approx&
\Omega^3 m_P.
\end{eqnarray}
The scale $\Omega$ has the value $10^{19}$--$10^{20}$. Here, $m_N$
and $l_N$ are the mass and radius of a nucleon, e.g. the proton.
The symbol $l_U$ denotes the observable radius of the universe,
that we define as the distance that light can travel in a Hubble
time $H_0^{-1}$. This time is roughly the age of our universe.
Finally, the mass of the universe $m_U$ is the energy contained in
a spatial region of radius $l_U$.

In fact, relations (\ref{lE}) and (\ref{mE}) are not independent.
For an elementary particle governed by quantum mechanics, the
typical effective size should be of the order of its Compton
wavelength, $l_N\approx \hbar/(cm_N)$. It therefore suffices to
explain, for instance, why $m_Pm_N^{-1}$ is of order $\Omega$.

Something similar happens with the scaling laws (\ref{lU}) and
(\ref{mU}). Assuming homogeneity and isotropy, $m_U$ is defined as
$4\pi l_U^3\rho^T_0/3$. Here, $\rho^T_0\equiv\rho_0+c^2
\Lambda/(8\pi G)$ is the total energy density. Hence, given the
relation between $l_U$ and $l_P$, formula (\ref{mU}) amounts to
the approximate equality $\rho_0^T\approx\rho_0^C$, where
$\rho_0^C\equiv 3 H_0^2/(8\pi G)$ is the critical density of a FRW
model at present. In a universe like ours, the scaling equation
for $m_U$ is thus a consequence of Eq. (\ref{lU}) and spatial
flatness.

Examining relations (\ref{lE})--(\ref{mU}), a length scale $l_S$
of order $\Omega^2$ in Planck units appears to be missing.
Roughly, this scale corresponds to the size of stellar
gravitational collapse determined by Chandrasekhar limit (or any
other similar mass limit) \cite{FPL}. Actually, for such
stellar-mass black holes, the formulas of the Schwarzschild radius
and the Chandrasekhar mass \cite{Weinberg} lead to
\begin{equation}\label{bh} l_S\approx \Omega^2 l_P,\hspace*{.8cm}
m_S\approx \Omega^2 m_P.\end{equation}

At this stage of our discussion, the only scaling laws that remain
unexplained are relations (\ref{mE}) and (\ref{lU}). In fact, one
of these approximate identities can be viewed as the definition of
$\Omega$, e.g. the equation for $l_U$. The appearance of large
numbers in our relations may then be understood, following Dirac
\cite{Dirac}, as a purely cosmological issue. Since $H_0^{-1}$ is
essentially the age of the universe, the fact that $\Omega\gg 1$
is just a consequence of the universe being so old. In addition,
it is easy to check that, given formula (\ref{lU}), the scaling
transformation for $m_N$ is equivalent to Eq. (\ref{rel}).
Therefore, the only coincidence of large numbers that needs
explanation is the Eddington-Weinberg relation.

Suppose now that nucleons (or hadronic particles in general) can
be described as elementary excitations of typical size $l_N$ in an
effective quantum theory. The number of physical degrees of
freedom in a spatial region of volume $V$ will be of the order of
$3V/(4\pi l_N^3)$. In a cosmological setting, it seems natural to
consider the Hubble radius as the largest size of the region in
which such an effective quantum description of particles may
exist, because it provides the scale of causal connection where
the microphysical interactions take place. For a homogeneous and
isotropic universe with negligible curvature, like the one we
inhabit, the FRW equations imply that $8\pi G\rho_0+c^2
\Lambda\approx 3 H_0^2$ \cite{Weinberg}. Given the positivity of
$\rho_0$, guaranteed by the dominant energy condition, the maximum
Hubble radius is thus close to $\sqrt{3/\Lambda}$. For an almost
flat FRW universe, the volume of the corresponding spatial region
is nearly $4\pi \sqrt{3/\Lambda^{3}}$. As a consequence, the
maximum number of observable degrees of freedom $N$ in this kind
of cosmological scenarios should roughly be
$\sqrt{27/(\Lambda^{3}l_N^{6})}$. Taking into account the
holographic $N$ bound (\ref{Nbound}), we then conclude
\begin{equation} \label{lrel}
l_N\approx (l_P^4 \Lambda^{-1})^{1/6}.\end{equation}

Using that $l_N m_N\approx l_Pm_P$, a relation that we have
already justified, we immediately obtain
\begin{equation}\label{mrel}m_N^3\approx
m_P^3(l_P^2\Lambda)^{1/2}.\end{equation} This approximate identity
reproduces Eq. (\ref{rel}) provided that the present Hubble radius
$cH_0^{-1}$ is close to $\Lambda^{-1/2}$. Therefore, the so-far
unexplained Eddington-Weinberg relation can be understood from a
holographic perspective, assuming an almost flat FRW cosmology, if
and only if the cosmological constant has a nearly dominant
contribution to the present energy density. This is ensured, e.g.,
by cosmic coincidence.

Note that the result $c^2\Lambda\approx H_0^2$ can be regarded as
a partial solution to the cosmological constant problems (the
value of $\Lambda$ and cosmic coincidence) in our (quasi)flat
universe if, adopting a different viewpoint, we take for granted
Bousso's proposal and Eq. (\ref{rel}). Alternatively, if we use
the Eddington-Weinberg relation and $c^2\Lambda\approx H_0^2$, the
arguments given above about the relation between $N$ and $l_N$
allow us to reach an approximate version of the $N$ bound for our
spacetime. Thus, we see that in a nearly homogeneous, isotropic
and flat universe like ours, the cosmological constant problems,
the $N$ bound, and the coincidence of large numbers are
interrelated.

In our application of the $N$ bound, we have argued that the
Hubble radius is the largest scale in which microphysics can act.
Nonetheless, our conclusions would not have changed if, as
proposed in Ref. \cite{BR} for cosmic holography, we had employed
the cosmological apparent horizon instead of the Hubble radius,
because they are approximately equal in quasiflat FRW models. We
have also made use of the fact that, for this kind of models, the
maximum Hubble radius is nearly $\sqrt{3/\Lambda}$ if $\Lambda$ is
positive. This is also the size of the cosmological horizon of the
de Sitter space with the same value of $\Lambda$. In (almost) flat
FRW cosmologies with a dominant $\Lambda$-term at late times, a
situation that apparently applies to our universe, any observer
has a future event horizon that tends asymptotically to such a de
Sitter horizon. Hence, our results would neither have been altered
had we replaced the maximum Hubble radius with the asymptotic
event horizon in all our considerations.

The fact that the $N$ bound provides an effective length scale for
microphysics, given by Eq. (\ref{lrel}), has played a central role
in our arguments. This fact has allowed us to understand the
origin of the Eddington-Weinberg relation. According to the
explanation that we have put forward, such a relation does not
hold at all times, but only when the cosmological constant
dominates the energy density. Although we expect this condition to
be satisfied at present and in the future, it excludes the early
stages of the evolution of the universe. In our theoretical
framework, the constants of nature $G$, $\hbar$, and $c$ do not
vary with time, and so we do not recover Dirac's cosmology
\cite{Dirac}.

In obtaining relation (\ref{lrel}), we have actually supposed that
the total number of degrees of freedom $N$ available in the
universe is roughly of the same order as the maximum number of
degrees observable in its baryonic content. It should be clear
that this assumption does not conflict with the fact that the
present energy density is not dominated by baryonic matter. More
importantly, since the number of baryonic degrees of freedom
cannot exceed $N$, the quantity $(l_P^4\Lambda^{-1})^{1/6}$
provides, in any case, a lower bound to the typical size of
nucleons $l_N$. Further discussion of this point will be presented
elsewhere.

The length scale (\ref{lrel}) has also been deduced by Ng,
although replacing $\Lambda^{-1}$ with the square of the
observable radius of the universe \cite{Ng}. However, he has
proposed to interpret $l_N$ as the minimum resolution length in
the presence of quantum gravitational fluctuations, instead of as
the typical size of particles in the effective QFT that describes
the baryonic content. From our viewpoint, this scale does not
provide a fundamental length limiting the resolution of spacetime
measurements, but rather restricts the number of degrees of
freedom available in the effective QFT. Concerning the value of
$l_N$, Ng proposes two ways to deduce it. In one of them, a
spatial region is considered as a Salecker-Wigner clock able to
discern distances larger than its Schwarzschild radius \cite{Ng}.
The question arises whether this interpretation is applicable to
the observable universe, because its Schwarzschild and Hubble
radii are of the same order of magnitude. The other line of
reasoning employs holographic arguments related to those presented
here. Nevertheless, since Ng uses the present size of the universe
instead of $\Lambda^{-1/2}$, it is not clear whether the
resolution scale that he obtains must be viewed as time
independent.

Let us return to expression (\ref{lU}) for the present Hubble
radius, which we have interpreted as the definition of $\Omega$.
We have argued that the fact that $\Omega\gg 1$ can be regarded as
a consequence of the old age of the universe, which is a
cosmological problem and not a numerical coincidence between
microscopic and macroscopic parameters. Nonetheless, using the $N$
bound and the present dominance of $\Lambda$, it is actually
possible to explain the appearance of the large scale $\Omega$
along very similar lines to those proposed by Banks for the
resolution of the cosmological constant problem \cite{Banks}. As
we have seen, when the energy density is nearly dominated by
$\Lambda$, the Hubble radius is close to $\sqrt{3/\Lambda}$. In
addition, the $N$ bound implies that this latter length is equal
to $l_P\sqrt{N\ln{2}/\pi}$. Recalling Eq. (\ref{lU}), we then
obtain
\begin{equation}\label{Om}\Omega\approx N^{1/6}.\end{equation} So,
$\Omega$ is a large number because our universe contains a huge
amount of degrees of freedom. From this perspective, the value of
$\Omega$ is fixed by $N$, which can be considered an input of the
theory that describes our world.

Finally, we want to present some brief comments about the entropy
of the universe. If the only entropic contribution were baryonic,
we could estimate it as $S_{b}\approx n_N$. Here, we have supposed
that each baryon has an associated entropy of order unity, and
$n_N$ is Eddington number, that can be calculated as the ratio of
the baryonic mass of the universe to the typical mass of a
nucleon. In a rough approximation (valid for our estimation of
orders of magnitude), we can identify the matter and the baryonic
energy densities. Taking into account cosmic coincidence, we can
then approximate $n_N$ by $m_Um_N^{-1}$. In this way, we get
$S_b\approx n_N\approx \Omega^4$. This is much less than the
maximum allowed entropy, which, from relation (\ref{Om}) and the
definition of $N$, is of the order of $\Omega^6$. An intermediate
entropic regime would be reached if the matter of the universe
collapsed into stellar-mass black holes. As we have commented,
this regime corresponds to the length scale $l_S\approx \Omega ^2
l_P$. One can check that, in this case, the entropy would be
$S_S\approx \Omega^5$. It is rather intriguing that $S_S$ matches
relatively well what seems to be the actual entropy of the
universe, $S_0$. The main contribution to this entropy comes from
super-massive black holes in galactic nuclei. Assuming that a
typical galaxy contains $10^{11}$--$10^{12}$ stellar masses $m_S$
and that the mass of its central black hole is $10^6$--$10^{7}$
$m_S$, it is straightforward to find that $S_0\approx 1$--$10^{3}$
$S_S$.

Summarizing, we have proved that, in the light of the holographic
principle, the relations between large numbers constructed from
microscopic and cosmological parameters are not independent of
other fine-tuning and coincidence problems that have a purely
cosmological nature. More explicitly, provided that the universe
can be approximately described by a spatially homogenous,
isotropic, and flat cosmological model and that the main
contribution to the present energy density comes from the
cosmological constant, it is possible to explain all the scaling
relations that motivated Dirac's large number hypothesis appealing
exclusively to basic principles and to the $N$ bound conjecture.

G.A.M.M. acknowledges DGESIC for financial support under Research
Project No. PB97-1218. S.C. was partially supported by CNPq. The
authors want to thank also L.J. Garay and P.F. Gonz\'{a}lez-D\'{\i}az for
helpful comments.


\begin{thebibliography}{}
\bibitem{entro} In fact, we call $S$ the entropy divided by
Boltzmann constant.
\end{thebibliography}
\end{document}